\documentclass[10pt,conference]{IEEEtran}

\usepackage{multirow}
\usepackage{graphicx}
\usepackage{url}
\usepackage{float}
\usepackage{array}
\usepackage{flushend}
\usepackage{amssymb}% http://ctan.org/pkg/amssymb
\usepackage{pifont}% http://ctan.org/pkg/pifont
\usepackage{amsmath}
\usepackage{capt-of}
\usepackage{color}
\newcommand\review[1]{\textcolor{black}{#1}}
\begin{document}

\title{Are SonarQube Rules Inducing Bugs?}
% \title{On the Fault Proneness of SonarQube Technical Debt Issues}

\author{\IEEEauthorblockN{Valentina Lenarduzzi}
\IEEEauthorblockA{Lahti-Lappeenranta University\\
Lahti-Lappeenranta, Finland\\
valentina.lenarduzzi@lut.fi}
\and
\IEEEauthorblockN{Francesco Lomio}
\IEEEauthorblockA{Tampere University\\
Tampere, Finland\\
francesco.lomio@tuni.fi}
\and
\IEEEauthorblockN{Heikki Huttunen}
\IEEEauthorblockA{Tampere University\\
Tampere, Finland\\
 heikki.huttunen@tuni.fi}
\and
\IEEEauthorblockN{Davide Taibi}
\IEEEauthorblockA{Tampere University\\
Tampere, Finland\\
 davide.taibi@tuni.fi}
}

% \author{\IEEEauthorblockN{Blinded Authors -- \textit{Blinded Affiliations}}
% }

% \author[1]{Valentina Lenarduzzi}
% \author[1]{Davide Taibi}
% \author[1]{Francesco Lomio}
% \author[1,2]{Luca Pellegrini}
% \author[1]{Heikki Huttunen}

% \authormark{Lenarduzzi et al.}

% \address[1]{\orgname{Tampere University of Technology}, \orgaddress{\state{Tampere}, \country{Finland}}}

% \address[2]{\orgname{Raiffeisen Verband Suedtirol}, \orgaddress{\state{Bozen-Bolzano}, \country{Italy}}}

% \corres{*Valentina Lenarduzzi,\email{valentina.lenarduzzi@tut.fi}}

% \presentaddress{Tampere University of Technology.\\Korkrankoulukatu 1\\33720 Tampere (Finland)}

\maketitle

\begin{abstract}
\textit{Background.} The popularity of tools for analyzing Technical Debt, and particularly the popularity of SonarQube, is increasing rapidly. SonarQube proposes a set of coding rules, which represent something wrong in the code that will soon be reflected in a fault or will increase maintenance effort. However, our local  companies were not confident in the usefulness of the rules proposed by SonarQube and contracted us to investigate the fault-proneness of these rules.   \\
\textit{Objective.} In this work we aim at understanding which SonarQube rules are actually  fault-prone and to understand which machine learning models can be adopted to accurately identify fault-prone rules.  \\
\textit{Method.} We designed and conducted an empirical study on 21 well-known mature open-source projects. We applied the SZZ algorithm to label the fault-inducing commits. We analyzed the fault-proneness by comparing the classification power of seven  machine learning models.  \\
\textit{Result.} Among the 202 rules defined for Java by SonarQube, only 25 can be considered to have relatively low fault-proneness. Moreover, violations considered as ''bugs'' by SonarQube were generally not fault-prone and, consequently, the  fault-prediction power of the model proposed by SonarQube is extremely low.\\
\textit{Conclusion.} The rules applied by SonarQube for calculating technical debt should be thoroughly investigated and their harmfulness needs to be further confirmed. Therefore, companies should carefully consider which rules they really need to apply, especially if their goal is to reduce fault-proneness.
\end{abstract}

% \keywords{Technical Debt, SonarQube, coding style, code smells, architectural smells, static analysis, machine learning}

% \jnlcitation{\cname{%
% \author{Williams K.}, 
% \author{B. Hoskins}, 
% \author{R. Lee}, 
% \author{G. Masato}, and 
% \author{T. Woollings}} (\cyear{2016}), 
% \ctitle{A regime analysis of Atlantic winter jet variability applied to evaluate HadGEM3-GC2}, \cjournal{Q.J.R. Meteorol. Soc.}, \cvol{2017;00:1--6}.}

\section{Introduction}
\label{Intro}

The popularity of tools for analyzing technical debt, such as SonarQube, is increasing rapidly. In particular, SonarQube has been adopted by more than 85K organizations~\footnote{https://www.sonarqube.org} including nearly 15K public open-source projects~\footnote{https://sonarcloud.io/explore/projects}. SonarQube analyzes code compliance against a set of rules. If the code violates a rule, SonarQube adds the time needed to refactor the violated rule as part of the technical debt. SonarQube also identifies a set of rules as ''bugs'', claiming that they ''represent something wrong in the code and will soon be reflected in a fault''; moreover, they also claim that zero false positives are expected from ''bugs''
\footnote{\label{sq-rules}SonarQube Rules: https://tinyurl.com/v7r8rqo}.

Four local companies \review{have been using SonarQube for more than five years to detect possible issue in their code,} reported that their developers do not believe that the rules classified as bugs can actually result in  faults. \review{ Moreover, they also reported that the manual customization of the SonarQube  out-of-the-box set of rules (named "the Sonar way"\footnote{SonarQube Quality Profiles: 
https://tinyurl.com/wkejmgr }) is very subjective and their developers did not manage to agree on a common set of rules that should be enforced.  Therefore, the companies  asked us to understand if it is possible to use machine learning to reduce the subjectivity of the customization of the SonarQube model, considering only rules that are actually fault-prone in their specific context. }

\review{SonarQube is not the most used static analysis tool on the market. Other tools such as Checkstyle, PMD and FindBugs are more used, especially in Open Source Projects~\cite{VassalloEMSE19} and in research~\cite{LenarduzziSEDA2018}. However, the adoption of another tool in the DevOps pipeline requires extra effort for the companies, including the training and the maintenance of the  tool itself. If the SonarQube rules  actually resulted fault-prone, our companies  would not need to invest extra effort to adopt and maintain other tools.}

% practitioners are reluctant to customize it and commonly rely on the "sonar way" standard rule-set~\cite{Vassallo2018}. Querying the SonarQube public instance APIs~\footnote{https://docs.sonarqube.org/display/DEV/API+Basics}, we can see that more than 98\% of the public projects (14,732 projects up to  14,957) uses the "sonar way" rule set. 
% However, even if developers are not sure about the usefulness of the rules ~\cite{Vassallo2018},~\cite{TaibiIST2017}, they do pay attention to different rules categories and priorities  and  remove violations related to rules with high severity~\cite{Vassallo2018} in order to avoid the possible risk of faults~\cite{TaibiIST2017}. Therefore,  investing effort for tasks that do not bring the expected benefits. 

At the best of our knowledge, not studies have investigated the fault-proneness of SonarQube rules, and therefore, we accepted the challenge and we designed and conducted this study. 
% Several studies, instead, have analyzed the impact of code smells ~\cite{Fowler1999} on faults~\cite{Yamashita2013}, \cite{Yamashita2014}, \cite{Deligiannis2004}\cite{Palomba19}. 
At best, only a limited number of studies have considered SonarQube rule violations~\cite{Falessi2017}, \cite{Tollin2017}, but they did not investigate the impact of the SonarQube violations considered as "bugs" on faults. 

% In this work, we aim to understand whether the presence of SonarQube rule violations is frequently associated with faults reported in issue tracking systems (Jira) and whether fault-fixing is also associated with the removal of the same violation. Moreover, in order
% to help practitioners understand which violations commonly generate faults, we aim at identifying a set of violations that have a higher correlation with faults in order to rank them based on their fault-proneness. Finally, we aim at analyzing the accuracy of the quality model provided by SonarQube in order to understand whether the rules it classifies as ''bugs'' are more fault-prone than other rule sets.

The goal of this work is twofold:
\begin{itemize}
\item Analyze the fault-proneness of  SonarQube rule violations, and in particular, understand if rules classified as ''bugs'' are more fault-prone than security and maintainability rules. 

\item Analyze the accuracy of the quality model provided by SonarQube in order to understand the fault-prediction accuracy of  the rules  classified as ''bugs''.

\end{itemize}

SonarQube and issue tracking systems adopt similar terms for different concepts. Therefore, in order to clarify the terminology adopted in this work, we define ~\textit{SQ-Violation} as a violated SonarQube rule that  generated a SonarQube "issue" and~\textit{fault} as an incorrect step, process, data definition, or any unexpected behavior in a computer program inserted by a developer, and reported by Jira issue-tracker.  We also use the term "fault-fixing" commit for commits where the developers have clearly reported bug fixing activity and "fault-inducing" commits for those commits that are responsible for the introduction of a fault.

The remainder of this paper is structured as follows. In Section~\ref{SonarQube} we introduce SonarQube and the SQ-Violations adopted in this work. In Section~\ref{Background} we present the background of this work, introducing the SonarQube violations and the different machine learning algorithms applied in this work. In Section~\ref{CS}, we describe the case study design.   
Section~\ref{Results} presents respectively the obtained results. Section~\ref{Threat} identifies threats to validity while  Section~\ref{relworks} describes related works.  Finally, conclusions are drawn in Section~\ref{Conclusions}. 
\section{Background}
\label{Background}

\subsection{SonarQube}
\label{SonarQube}
SonarQube is one of the most common Open Source static code analysis tools adopted both in academia~\cite{LenarduzziICSE2017},\cite{LenarduzziSEDA2018} and in industry~\cite{VassalloEMSE19}. SonarQube is provided as a service from the sonarcloud.io platform or it can be downloaded and executed on a private server.

SonarQube calculates several metrics such as the number of lines of code and the code complexity, and verifies the code's compliance against a specific set of "coding rules" defined for most common development languages. 
% Moreover, it defines a set of thresholds ("quality gates") for each metric and rule. 
In case the analyzed source code violates a coding rule or if a metric is outside a predefined threshold, SonarQube generates an "issue". 
% The time needed to remove these issues (remediation effort) is used to calculate the remediation cost and the technical debt.
SonarQube includes Reliability, Maintainability and Security rules. 
% Moreover, SonarQube claims that zero false positives are expected from the Reliability and Maintainability rules while there could be some false positives \textsuperscript{\ref{sq-rules}}.

Reliability rules, also named "bugs" create issues (code violations) that ''represents something wrong in the code'' and that will soon be reflected in a bug.  ''Code smells'' are considered  ''maintainability-related issues'' in the code that decreases code readability and code modifiability. It is important to note that the term ''code smells'' adopted in SonarQube does not refer to the commonly known code smells defined by Fowler et al.~\cite{Fowler1999} but to a  different set of rules. Fowler et al.~\cite{Fowler1999} consider code smells as ''surface indication that usually corresponds to a deeper problem in the system'' but they can be indicators of different problems (e.g., bugs,  maintenance effort, and code readability) while rules classified by SonarQube as ''Code Smells'' are only referred to maintenance issues. Moreover,  only four of the 22 smells proposed my Fowler et al. are included in the rules  classified as ''Code Smells'' by SonarQube (Duplicated Code, Long Method, Large Class, and Long Parameter List).   

SonarQube also  classifies the rules into five \textit{severity} levels\footnote{SonarQube Issues and Rules Severity:'  https://docs.sonarqube.org/display/SONAR/Issues Last Access:May 2018}: Blocker, Critical, Major, Minor, and Info. 
% \begin{itemize}
% \item \textit{BLOCKER}: ''Bug with a high probability to impact the behavior of the application in production: memory leak, unclosed JDBC connection.'' SonarQube recommends to immediately review this issue
% \item \textit{CRITICAL}: ''Either a bug with a low probability to impact the behavior of the application in production or an issue which represents a security flaw: empty catch block, SQL injection'' SonarQube recommends to immediately review this issue 
% \item \textit{MAJOR}: ''Quality flaw which can highly impact the developer productivity: uncovered piece of code, duplicated blocks, unused parameters''
% \item \textit{MINOR}: ''Quality flaw which can slightly impact the developer productivity: lines should not be too long, switch statements should have at least 3 cases, ...''
% \item \textit{INFO}: ''Neither a bug nor a quality flaw, just a finding.''
% \end{itemize}

In this work, we focus on the sq-violations, which are reliability rules classified  as ''bugs'' by SonarQube, as we are interested in understanding whether they are related to faults. 

SonarQube includes more than 200 rules for Java (Version 6.4). In the replication package (Section~\ref{sec:Replicability}) we report all the violations present in our dataset. In the remainder of this paper, column ''\textit{squid}'' represents the original rule-id (SonarQube ID) defined by SonarQube. We did not rename it, to ease the replicability of this work.  In the remainder of this work, we will refer to the different sq-violations with their id (squid).
The complete list of violations can be found in the file "SonarQube-rules.xsls" in the online raw data.

\subsection{Machine Learning Techniques}\label{MachineLearning}
In this Section, we describe the machine learning techniques adopted in this work to predict the fault-proneness of sq-violations. Due to the nature of the task, all the models used for this work were used for classification. We compared eight machine learning models. Among these, we used a generalized linear model: Logistic Regression \cite{Cox1958}; one tree based classifier: Decision Tree \cite{Breiman1984ClassificationTrees}; and 6  \textit{ensemble classifiers}: Bagging \cite{Breiman1996BaggingPredictors}, Random Forest \cite{breiman2001random}, Extremely Randomized Trees \cite{Geurts2006ExtremelyTrees}, AdaBoost \cite{Freund1997ABoosting}, Gradient Boosting \cite{FriedmanGreedyMachine}, and XGBoost \cite{Chen2016XGBoost:Systemb} which is an optimized implementation of Gradient Boosting.
\review{All the models, except the XGBoost, were implemented using the library \textit{Scikit-Learn}\footnote{https://scikit-learn.org}, applying the default parameters for building the models. For the ensamble classifiers we alwasys used 100 estimators. The XGBoost classifier was implemented using the \textit{XGBoost} library\footnote{https://xgboost.readthedocs.io} also trained with 100 estimators.}

\subsubsection{Logistic Regression~\cite{Cox1958}}
Contrary to the linear regression, which is used to predict a numerical value, Logistic Regression is used for predicting the category of a sample. Particularly, a binary Logistic Regression model is used to estimate the probability of a binary result ($0$ or $1$) given a set of independent variables. Once the probabilities are known, these can be used to classify the inputs in one of the two classes, based on their probability to belong to either of the two. 

Like all linear classifiers, Logistic Regression projects the $P$-dimensional input $\bf x$ into a scalar by a dot product of the learned weight vector ${\bf w}$ and the input sample: ${\bf w} \cdot {\bf x} + w_0$, where $w_0\in\mathbb{R}$ the constant intercept.
To have a result which can be interpreted as a class membership probability---a number between $0$ and $1$---Logistic Regression passes the projected scalar through the logistic function (sigmoid). This  function, for any given input $x$, returns an output value between $0$ and $1$. 
The logistic function is defined as
 \[\sigma(x) = \frac{1}{1+e^{-x}}.\]
 Where the class probability of a sample ${\bf x} \in \mathbb{R}^P$ is modeled as 
 \[Pr(c=1 \mid {\bf x}) = \frac{1}{1+e^{-({\bf w} \cdot {\bf x} + w_0)}}.\]
Logistic Regression is trained through maximum likelihood: the model's parameters are estimated in a way to maximize the likelihood of observing the inputs with respect to the parameters $\bf w$ and $w_0$.
We chose to use this model as baseline as it requires limited computational resources and it is easy to implement and fast to train.% Moreover, it doesn't need the inputs to be scaled nor it needs to be tuned.

\subsubsection{Decision Tree Classifier~\cite{Breiman1984ClassificationTrees}}
Utilizes a decision tree to return an output given a series of input variables. Its tree structure is characterized by a \textit{root node} and multiple \textit{internal nodes}, which are represented by the input variable, and \textit{leaf}, corresponding to the output. The nodes are linked between one another through branches, representing a test. The output is given by the decision path taken.
A decision tree is structured as a if-then-else diagram: in this structure, given the value of the variable in the root node, it can lead to subsequent nodes through branches following the result of a test. This process is iterated for all the input variables (one for each node) until it reaches the output, represented by the leaves of the tree.

In order to create the best structure, assigning each input variable to a different node, a series of metrics can be used. Amongst these we can find the \textit{GINI impurity} and the \textit{information gain}:
\begin{itemize}
\item Gini impurity measures how many times randomly chosen inputs would be wrongly classified if assigned to a randomly chosen class;
\item Information gain measures how important is the information obtained at each node related to its outcome: the more important is the information obtained in one node, the purer will be the split.
\end{itemize}

In our models we used the Gini impurity measure to generate the tree as it is more computationally efficient.
% \review{The model selects the features to be used for building the trees. Since all the models used in this study are three-based, we applied this approach for all the classifiers.} 
The reasons behind the choice of decision tree models and Logistic Regression, are their simplicity and easy implementation. Moreover, the data does not need to be normalized, and the structure of the tree can be easily visualized.
However, this model is prone to overfitting, and therefore it cannot generalize the data. Furthermore, it does not perform well with imbalanced data, as it generates a biased structure.

\subsubsection{Random Forest~\cite{breiman2001random}} is an ensemble technique that helps to overcome overfitting issues of  the decision tree. The term ensemble indicates that these models use a set of simpler models to solve the assigned task. In this case, Random Forest uses an ensemble of decision trees.

An arbitrary number of decision trees is generated considering a randomly chosen subset of the samples of the original dataset~\cite{Breiman1996BaggingPredictors}. This subset is created with replacement, hence a sample can appear multiple times. Moreover, in order to reduce the correlation between the individual decision trees a random subset of the features of the original dataset. In this case, the subset is created without replacement.
Each tree is therefore trained on its subset of the data, and it is able to give a prediction on new unseen data. The Random Forest classifier uses the results of all these trees and averages them to assign a label to the input.
By randomly generating multiple decision trees, and averaging their results, the Random Forest classifier is able to better generalize the data. Moreover, using the random subspace method, the individual trees are not correlated between one another. This is particularly important when dealing with a dataset with many features, as the probability of them being correlated between each other increases.

\subsubsection{Bagging~\cite{Breiman1996BaggingPredictors}} Exactly like the Random Forest model, the Bagging classifier is applied to an arbitrary number of decision trees which are constructed choosing a subset of the samples of the original dataset. 
The difference with the Random Forest classifier is in the way in which the split point is decided: while in the Random Forest algorithm the splitting point is decided base on a random subset of the variables, 
the Bagging algorithm is allowed to look at the full set of variable to find the point minimizing the error. This translates in structural similarities between the trees which do not resolve the overfitting problem related to the single decision tree.
This model was included as a mean of comparison with newer and better performing models.

\subsubsection{Extremely Randomized Trees~\cite{Geurts2006ExtremelyTrees}}
 (ExtraTrees)~\cite{Geurts2006ExtremelyTrees}, provides a further randomization degree to the Random Forest.
For the Random Forest model, the individual trees are created by randomly choosing subsets of the dataset features. In the ExtraTrees model the way each node in the individual decision trees are split is also randomized. Instead of using the metrics seen before to find the optimal split for each node (Gini impurity and Information gain), the cut-off choice for each node is completely randomized, and the resulting splitting rule is decided based on the best random split.
Due to its characteristics, especially related to the way the splits are made at the node level, the ExtraTrees model is less computationally expensive than the Random Forest model, while retaining a higher generalization capability compared to the single decision trees.

\subsubsection{AdaBoost~\cite{Freund1997ABoosting}}
is another ensemble algorithm based on \textit{boosting}~\cite{Schapire1990TheLearnability}  where the individual decision trees are grown sequentially. Moreover, a weight is assigned to each sample of the training set.
Initially, all the samples are assigned the same weight. The model trains the first tree in order to minimize the classification error, and after the training is over, it increases the weights to those samples in the training set which were misclassified. Moreover, it grows another tree and the whole model is trained again with the new weights. This whole process continues until a predefined number of trees has been generated or the accuracy of the model cannot  be improved anymore.
Due to the many decision trees, as for the other ensemble algorithms, AdaBoost is less prone to overfitting and can, therefore, generalize better the data. Moreover, it automatically selects the most important features for the task it is trying to solve. However, it can be more susceptible to the presence of noise and outliers in the data.

\subsubsection{Gradient Boosting~\cite{FriedmanGreedyMachine}}
also uses an ensemble of individual decision trees which are generated sequentially, like for the AdaBoost.
The Gradient Boosting trains at first only one decision tree and, after each iteration, grows a new tree in order to minimize the loss function. Similarly to the AdaBoost, the process stops when the predefined number of trees has been created or when the loss function no longer improves.

\subsubsection{XGBoost~\cite{Chen2016XGBoost:Systemb}} can be viewed as a better performing implementation of the Gradient Boosting algorithm, as it allows for faster computation and parallelization. For this reason it can yield better performance compared to the latter, and can be more easily scaled for the use with high dimensional data.
\section{Case Study Design}
\label{CS}
We designed our empirical study as a case study based on the guidelines defined by Runeson and H\''{o}st~\cite{Runeson2009}.
In this Section, we describe the empirical study including the goal and the research questions, the study context, the data collection and the data analysis. 

\subsection{Goal and Research Questions}
As reported in Section 1, our goals are to analyze the fault-proneness of  SonarQube rule violations (SQ-Violations) and  the accuracy of the quality model provided by SonarQube. Based on the aforementioned goals, we derived the following three research questions (\textbf{RQs}).

\vspace{2mm}
\begin{tabular}
{@{}p{0.5cm}p{7.5cm}@{}}
\textbf{RQ1} & \textbf{Which are the most fault-prone SQ-Violations? }\\ & In this RQ, we aim to understand whether the introduction of a set of SQ-Violations is correlated with the introduction of faults in the same commit and to prioritize the SQ-Violations based on their fault-proneness. \\ &Our hypothesis is that  a set of SQ-Violations should be responsible for the introduction of bugs.\\
\textbf{RQ2} &  \textbf{Are SQ-Violations classified as ''bugs'' by SonarQube more fault-prone than other rules?} \\ & Our hypothesis is that reliability rules (''bugs'') should be more fault-prone that   maintainability rules (''code smells'') and security rules.\\
\textbf{RQ3} & \textbf{What is the fault prediction accuracy of the SonarQube quality model based on violations classified as ''bugs''? }\\ & SonarQube claims that whenever a violation is classified as a ''bug'', a fault will develop in the software. \\ & Therefore, we aim at analyzing the fault prediction accuracy of the rules that are classified as ''bugs'' by measuring their precision and recall.\\
\end{tabular}

\subsection{Study Context}
\label{Context}

In agreement with the four companies, we considered  open source projects available in the Technical Debt Dataset~\cite{LenarduzziPromise2019}.
\review{The reason for considering open source projects instead of their private projects is that not all the companies would have allowed us to perform an historical analysis of all their commits. Moreover, with closed source projects the whole process cannot be replicated and verified  transparently.} 

For this purpose, the four companies selected together 21 out of 31 projects available, based on the ones that were more similar to their internal projects considering similar \review{ project age,  size, usage of patterns used and other} criteria that we cannot report for reason of NDA. 

% The dataset includes widely used Java projects from the Apache Software Foundation (ASF) repository\footnote{http://apache.org}. 
% The  projects can be considered industrial and mature, due the strict review process required by the ASF. 
% Moreover, the included projects have to keep on reviewing their code and follow a strict quality process\footnote{https://incubator.apache.org/policy/process.html}. 

% The projects were 
% selected with a ''criterion sampling''\cite{Patton2002}, fulfilling all the following criteria: 

% \begin{itemize}
% \item Developed in Java
% \item Older than three years
% \item More than 500 commits
% \item More than 100 classes
% \item Using Github for source code versioning
% \item Projects reporting faults in Jira
% \item Projects where, in case of fault-fixing activities, developers report the fault-id in the commit message  
% \end{itemize}

% Moreover, as recommended by Nagappan et al. \cite{Nagappan2013}, we also tried to maximize diversity and representativeness, considering a comparable number of projects with respect to project age (number of years from the project creation), size (number of LOC of the last version), and application type (eg. web server, library, IDE, ...). 

% All the selected projects are available in a Git repository, and track their issues with Jira\footnote{https://issues.apache.org/jira/}.

The dataset includes the analysis of each commit of the projects from their first commit until the end of 2015 with SonarQube, information on all the Jira issues, and a classification of the  fault-inducing commits performed with the SZZ algorithm~\cite{SZZ}. 

% Since faults are commonly discovered after the code has already been committed, we 

% The dataset analyzed the projects from their first commit until the end of 2015, considering all the faults raised until the end of March 2018. Therefore, we can ensure that the vast majority of the faults introduced in the commits should have been discovered after more than two years.  

In Table~\ref{tab:SelectedProjects}, we report the list of  projects we considered together with the number of analyzed commits, the project size (LOC) of the last analyzed commits, the number of faults identified in the selected commits, and the total number of SQ-Violations.

% \todo [inline] {@Valentina. update the project commits and raw data with the latest data analyzed}
\begin{table}
\footnotesize
\centering
\caption{The selected projects} 
\label{tab:SelectedProjects} 
\begin{tabular}
{@{}p{1.6cm}|p{1.3cm}|p{1cm}|p{1cm}|p{1.5cm}@{}}
\hline 
\textbf{Project Name} & \textbf{Analyzed commits}& \textbf{Last commit LOC}& \textbf{Faults}& \textbf{SonarQube Violations}\\ \hline 
Ambari & 9727 & 396775 & 3005 & 42348\\ \hline 
Bcel & 1255 & 75155 & 41 & 8420 \\ \hline 
Beanutils &1155 & 72137 & 64 & 5156\\ \hline 
Cli & 861 & 12045 & 59 & 37336\\ \hline 
Codec & 1644 & 34716 & 57 & 2002\\ \hline 
Collections & 2847 & 119208 & 103 & 11120\\ \hline 
Configuration & 2822 & 124892 & 153 & 5598\\ \hline 
Dbcp & 1564 & 32649 & 100 & 3600\\ \hline 
Dbutils & 620 & 15114 & 21 & 642\\ \hline 
Deamon & 886 & 3302 & 4 & 393\\ \hline 
Digester & 2132 & 43177 & 23 & 4945\\ \hline 
FileUpload & 898 & 10577 & 30 & 767\\ \hline 
Io & 1978 & 56010 & 110 & 4097\\ \hline 
Jelly & 1914 & 63840 & 45 & 5057\\ \hline 
Jexl &1499 & 36652 & 58 & 34802 \\ \hline 
Jxpath & 596 & 40360 & 43 & 4951 \\ \hline 
Net & 2078 & 60049 & 160 & 41340\\ \hline 
Ognl & 608 & 35085 & 15 & 4945\\ \hline 
Sshd & 1175 & 139502 & 222 & 8282\\ \hline 
Validator &1325 & 33127 & 63 & 2048 \\ \hline 
Vfs & 1939 & 59948 & 129 & 3604\\ 
\hline
\textbf{Sum} & \textbf{39.518} & \textbf{1,464,320}& \textbf{4,505}& \textbf{231,453}\\ 
\hline
\end{tabular}
\end{table}

\subsection{Data Analysis}
\label{DataAnalysis}
Before answering our RQs, we first executed the eight machine learning (ML) models, we compared their accuracy, and finally performed the residual analysis.

\begin{figure}
\centering
\includegraphics[width=\linewidth]{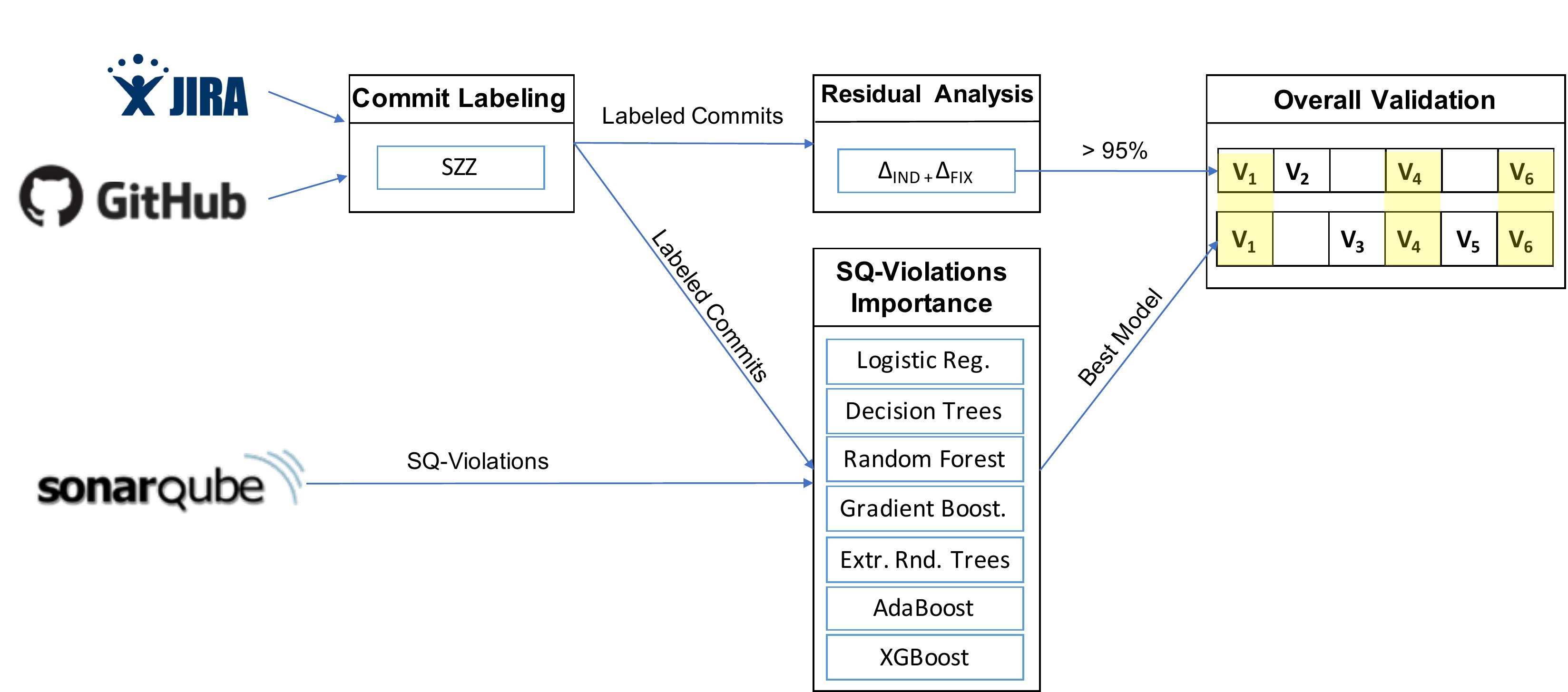}
\caption{The Data Analysis Process}
\label{fig:process}
\end{figure}

The next subsections describe the analysis process in details as depicted in Figure~\ref{fig:process}. 

\subsubsection{Machine Learning Execution} 
In this step we aim at comparing  fault-proneness prediction power of SQ-Violations by applying the eight machine learning models described in Section \ref{MachineLearning}.
% in order to confirm or reject the  results obtained in the residual analysis. 

Therefore we aim at predicting the fault-proneness of a commit (labeled with the SZZ algorithm) by means of the SQ-Violations introduced in the same commit. 
We used the SQ-Violations introduced in each commits as independent variables (predictors)  to determine if a commit is fault-inducing (dependent variable).

  %\todo [inline] { Here we aim at comparing the seven ML techniques with the goal of analyzing the importance of SQviolations}

%\todo [inline] { introduce how we use the techinques}
% In order to assess the importance of each SQ-Violation, we used the Random Forest classifier \cite{breiman2001random}. [1].	Random Forest has the ability to rank the input features based on their relative importance for prediction accuracy. This is important in this application, as we wish to identify those SQ-Violations that are linked to fault-inducing commits.
% Next, we applied Recursive Feature Elimination (RFE) to select and reduce the relevant set of SQ-Violations (RQ2).

% As a confirmation of the feature selection approach, we used $\ell_1$-regularized logistic regression model \cite{friedman2010regularization}. With $\ell_1$ regularization, this model inherently selects a good set of input features and discards non-important features altogether. Note that this approach results in a fixed set of features (without an internal ranking), while random forest results in a list of ranked features. Thus, the two approaches are complementary to each other.
After training the eight models described in Section~\ref{MachineLearning}, we performed a second analysis retraining the models using a \textit{drop-column mechanism}~\cite{drop-col}. This mechanism is a simplified variant of the exhaustive search~\cite{yoon2005feature}, which iteratively tests every subset of features for their classification performance. The full exhaustive search is very time-consuming requiring $2^P$ train-evaluation steps for a $P$-dimensional feature space. Instead, we look only at dropping individual features one at a time, instead of all possible groups of features.
%\todo [inline] {@Heikki, is there a better reference, or may we add this link as ref.?}
%\todo [inline] {I think this is the best ref I can find. Drop column and other exhaustive search variants are commonly known but there does not seem to be a systematic study elsewhere. ---Heikki}

%Each model used the SQ-Violation as input features and as output returned whether the commit is fault-inducing or not. 

More specifically,  a model is trained $P$ times, where $P$ is the number of features, iteratively removing one feature at a time, from the first to the last of the dataset. The difference in cross-validated test accuracy between the newly trained model and the baseline model (the one trained with the full set of features) defines the importance of that specific feature. The more the accuracy of the model drops, the more important for the classification is the specific feature. 

The feature importance of the SQ-Violation has been calculated for all the machine learning models described, but we considered only the importance calculated by the most accurate model (cross-validated with all $P$ features, as described in the next section), as the feature importances of a poor classifier are likely to be less reliable.

%\todo [inline] {describe the drop column}
%  \todo [inline] {Do we still need this third method (RFE) too? LR will now answer to RQ2 already / Heikki}

% \vspace{-2mm}
\subsubsection{Accuracy Comparison}
\label{sec:accuracy}
Apart from ranking the SQ-Violations by their importance, we first need to confirm the validity of the prediction model. If the predictions obtained from the ML techniques are not accurate, the feature ranking would also become questionable. To assess the prediction accuracy, we performed a 10-fold cross-validation, dividing the data in 10 parts, \textit{i.e.,} we trained the models ten times always using 1/10 of the data as a testing fold. For each fold, we evaluated the classifiers by calculating a number of accuracy metrics (see below). 
The data related to each project have been split in 10 sequential parts, thus respecting the temporal order, and the proportion of data for each project. The models have been trained iteratively on group of data preceding the test set. The temporal order was also respected for the groups included in the training set: as an example, in fold 1 we used group 1 for training and group 2 for testing, in fold 2 groups 1 and 2 were used for training and group 3 for testing, and so on for the remaining folds.  
%\todo[inline]{explain how the cv works}
% \todo [inline]{This following part must be modified, or maybe removed? / Francesco}
%  As a second part of our pipeline, we analyze the results of the feature selection by looking at both models: The result of training the random forest is a vector of SQ-Violations with their importance in relation with  faults; and the result of training the logistic regression model is a subset of SQ-Violations, that should contain the highest prediction power regarding the fault type.
%  \todo [inline] {A figure would help to understand ''our pipeline'' / Outi}
%  \todo [inline] {True. I added thje figure. The figure also helped to better structure the analysis section. I splitted the overal analysis in two subsections adding  a ''residual analysis'' section }

%\todo [inline] { Report why we do not use precision and recall (number of classes not balanced) }

As accuracy metrics, we first calculated precision and recall. However, as suggested by ~\cite{Powers2011}, 
these two measures present some biases as they are mainly focused on positive examples and predictions and they do not capture any information about the rates and kind of errors made. 

The contingency matrix (also named confusion matrix), and the related f-measure help to overcome this issue. Moreover, as recommended by~\cite{Powers2011}, the Matthews Correlation Coefficient (MCC) should be also considered to understand possible  disagreement between actual values and predictions as it involves all the four quadrants of the contingency matrix.

From the contingency matrix, we retrieved the measure of \textit{true negative rate} (TNR), which measures the percentage of negative sample correctly categorized as negative, \textit{false positive rate} (FPR) which measures the percentage of negative sample misclassified as positive, and \textit{false negative rate} (FNR), measuring the percentage of positive samples misclassified as negative. The measure of \textit{true positive rate} is left out as equivalent to the recall.
The way these measures were calculated can be found in Table \ref{tab:formula}.

\begin{table}[!ht]
\centering
\caption{Accuracy Metrics Formulae}
\label{tab:formula}
\begin{tabular}{c|c}
\hline
\textbf{Accuracy Measure} & \textbf{Formula}\\
\hline
Precision &\(\frac{TP}{FP + TP}\) \\ \hline
Recall &\(\frac{TP}{FN + TP}\)\\ \hline
MCC &\(\frac{TP * TN - FP * FN}{\sqrt{(FP + TP)(FN + TP)(FP + TN)(FN + TN)}}\)\\ \hline
f-measure &\(2* \frac{precision * recall}{precision + recall}\)\\ \hline
TNR &\(\frac{TN}{FP + TNe}\)\\ \hline
FPR &\(\frac{FP}{TN + FP}\)\\ \hline
FNR &\(\frac{FN}{FN + TP}\)\\ \hline

\end{tabular}
\smallskip

TP: True Positive; TN: True Negative; FP: False Positive; FN: False Negative
\vspace{-4mm}
\end{table}

Finally, to graphically compare the true positive  and the false positive rates, 
we calculated the Receiver Operating Characteristics (ROC), and the related Area Under the Receiver Operating Characteristic Curve (AUC): the probability that a classifier will rank a randomly chosen positive instance higher than a randomly chosen negative one.

% Moreover, from the contingency matrix we calculated the chi squared test, so as to obtain the p-value. 
% We accept results for p-valuer lower than 0.05 as customary done in empirical software engineering. 

% \todo[inline]{The list of metrics looks super long now. Should we shorten it? I'm explaining some reasons below, but I would shorten the above list anyway. / Heikki}
% \todo[inline]{I agree, but I would expect that removing one of them we will have a negative comment from some reviewer that probably love the missing one ;-)}

In our dataset, the proportion of the two types of commits is not even: a large majority (approx. 90 \%) of the commits were non-fault-inducing, and a plain accuracy score would reach high values simply by always predicting the majority class. On the other hand, the ROC curve (as well as the precision and recall scores) are informative even in seriously unbalanced situations.

% Therefore, we cannot rely on precision, recall, MCC and f-measure, as also suggested by ~\cite{Powers2011}. The ROC curve ad the related AUC are more  informative even in seriously unbalanced situations. \\

% \begin{figure}[H]
% \centering
% \includegraphics[width=0.9\linewidth]{./Figure/Picture1.png}
% \end{figure}

% \begin{figure}[H]
% \centering
% \includegraphics[width=0.4\linewidth]{./Figure/Picture2.png}
% \end{figure}

\subsubsection{SQ-Violations Residual Analysis}
The results from the previous ML techniques show a set of SQ-Violations related with fault-inducing commits. However, the relations obtained in the previous analysis do not imply causation between faults and SQ-Violations. 
% This step helps to get more accurate results on the correlations.

In this step, we analyze which violations were introduced in the fault-inducing commits and then removed in the fault-fixing commits. We performed this comparison at the file level. Moreover, we did not consider cases where the same violation was introduced in the fault-inducing commit, removed, re-introduced in commits not related to the same fault, and finally removed again during the fault-fixing commit.

In order to understand which SQ-Violations were introduced in the fault-inducing commits (IND) and then removed in the fault-fixing  commit (FIX), we  analyzed the residuals of each SQ-Violation by calculating: 
\vspace{-2mm}
\begin{equation*}
Residual = \Delta_{IND} + \Delta_{FIX}
\end{equation*}
% \begin{equation*}
% Residual = (\#SQ-Violations Introduced in the Fault-Inducing commit)-(\#SQ-Violations Removed in the Fault-Fixing commit) \end{equation*}
where $\Delta_{IND}$ and $\Delta_{FIX}$ are calculated as:
% \begin{itemize}
% \item [$\Delta_{IND}$] = $\Delta_{SQ-Violations (IND-1,IND)}$
% \item [$\Delta_{FIX}$] = $\Delta_{SQ-Violations (IND,FIX)}$
% \end{itemize}

\begin{center}
 $\Delta_{IND}$ = \textit{\#SQ-Violations introduced in the fault-inducing commit}
%  $\Delta_{SQ-Violations (IND-1,IND)}$

$\Delta_{FIX}$ =\textit{\#SQ-Violations removed in the fault-fixing commit}
% $\Delta_{SQ-Violations (IND,FIX)}$
\end{center}

% The residual represents the number of violations introduced in the fault-inducing commits ()

\begin{figure}
    \centering
    \includegraphics[width=1.016\linewidth]{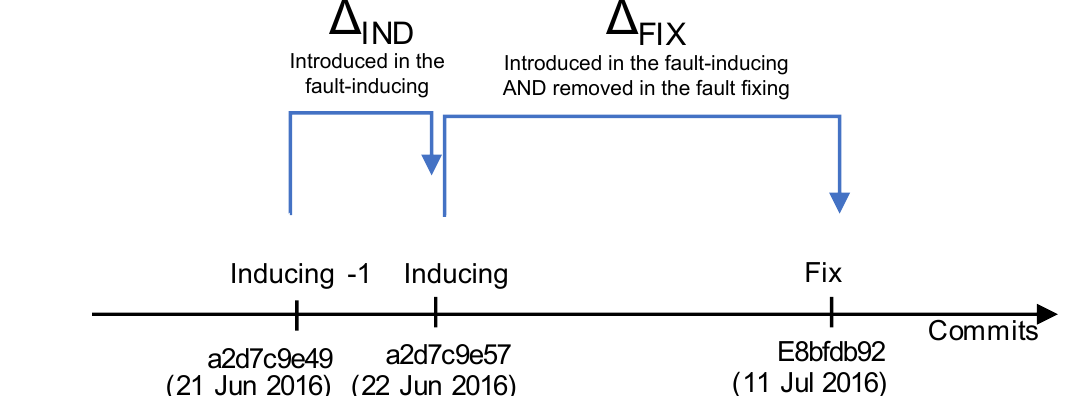}
    \caption{Residuals Analysis} \label{fig:DeltaAnalysis}
    \vspace{-2mm}
\end{figure}

% Figure \ref{fig:sets} shows the possible cases of  introduced and removed violations while 
Figure~\ref{fig:DeltaAnalysis} schematizes the residual analysis.

% \begin{table*}[!h]
% \centering
% \begin{tabular}{ll}

% \begin{minipage}{0.48\textwidth}
% \includegraphics[width=1\linewidth]{./Figure/sets}
% \captionof{figure}{Possible cases of  introduced and removed violations} \label{fig:sets}
% \end{minipage}

% & 

% \begin{minipage}{0.48\textwidth}
% \includegraphics[width=1.016\linewidth]{./Figure/Delta.pdf}
% \captionof{figure}{Residuals Analysis} \label{fig:DeltaAnalysis}
% \end{minipage} \\
% \end{tabular}

% \end{table*}

\review{We calculated the residuals for each commit/fix pair, verifying the introduction of the SQ-Violation $V_{i}$ in the fault-inducing commit (IND) and the removal of the violation in the fault-fixing commit (FIX)}. If $\Delta_{IND}$ was lower than zero, no SQ-Violations were introduced in the fault-inducing commit. Therefore, we tagged such a commit as not related to faults. 

For each violation, the analysis of the residuals led us to two groups of commits: 
% (Figure \ref{fig:sets}): 
\begin{itemize}
\item \textit{Residual $>$ 0}: The SQ-Violations introduced in the fault-inducing commits were not removed during the fault-fixing. 
\item \textit{Residual $\leq$ 0}: All the SQ-Violations introduced in the fault-inducing commits were removed during the fault-fixing. If Residual $<$ 0, other SQ-Violations of the same type already present in the code before the bug-inducing commit were also removed. 
\end{itemize}

% The residual analysis process is depicted in Figure~\ref{fig:residuals}.

% \begin{figure}[H]
% \centering
% \includegraphics[width=0.4\linewidth]{./Figure/ResidualAnalysis}
% \caption{Residual Analysis Process}
% \label{fig:residuals}
% \end{figure}

For each SQ-Violations, we calculated descriptive statistics so as to understand the distribution of residuals. 

Then, we calculated the residual sum of squares (RSS) as:
\vspace{-1mm}
\begin{equation*}
RSS = \displaystyle \sum (Residual)^2
\end{equation*}

We calculated the percentage of residuals equal to zero as:
\begin{equation*}
\frac{\#zero\_residuals}{\#residuals}*100\%
\end{equation*}

Based on the residual analysis, we can consider violations where the percentage of zero residuals was higher than 95\% as a valid result.

\subsubsection{ RQ1: Which are the most fault-prone SQ-Violations?}
In order to analyze RQ1, we combined the results obtained from the best ML technique and from the residual analysis.
Therefore, if a violation has a high correlation with faults  but the percentage of the residual is very low, we can discard it from our model, since it will be valuable only in a limited number of cases.
As we cannot claim a cause-effect relationship without a controlled experiment, the results of the residual analysis are a step towards the identification of this relationship and the reduction of spurious correlations.

\subsubsection{RQ2: Are  SQ-Violations  classified  as  ”bugs”  by  Sonar-Qube  more  fault-prone  than  other  rules?}
The comparison of rules classified as ''bugs'' with other rules has been performed considering the results of the best ML techniques and the residual analysis, comparing the number of violations classified as ''bug'' that resulted to be fault-prone from RQ1. We expect bugs to be in the most faults-prone rules. 

\subsubsection{RQ3: What   is   the   fault   prediction   accuracy   of   the SonarQube   quality   model   based   on   violations classified  as  ''bugs''}
Since SonarQube considers every SQ-Violation tagged as a ''bug'' as ''something wrong in the code that will soon be reflected in a bug'', we also analyzed the accuracy of the model provided by SonarQube.

In order to answer our RQ3, we calculated the percentage of SQ-Violations classified as ''bugs'' that resulted in being highly fault-prone according to the previous analysis. 
Moreover, we also  analyzed the accuracy of the model calculating all the accuracy measures reported in Section~\ref{sec:accuracy}.

% We first labeled every commit, considering as ''sq-faulty'' every commit where a ''bug'' violation was introduced. Then we analyzed the accuracy of the model calculated the contingency matrix, precision and recall, and the Mathews correlation coefficient to compare the results with the commit labeled as fault-inducing by the SZZ. 

\subsection{Replicability}
\label{sec:Replicability}
In order to allow the replication of our study, we published the raw data in the replication package~\footnote{Replication Package: https://figshare.com/s/fe5d04e39cb74d6f20dd}.

\section{Results}
\label{Results}
In this work, we considered more than 37 billion effective lines of code and retrieved a total of 1,464,320 violations from 39,518 commits scanned with SonarQube. Table 1 reports the list of projects together with the number of analyzed commits and the size (in Lines of Code) of the latest analyzed commit. We retrieved a total of 4,505 faults reported in the issue trackers.

% The data collection required three months of computation time on a Linux Ubuntu server with 15 cores and 64 GB RAM, while the data analysis required 6 days on the same machine.

All the 202 rules available in SonarQube for Java were found in the analyzed projects. For reasons of space limitations, we will refer to the SQ-Violations only with their SonarQube id number (SQUID). The complete list of  rules, together with their description is reported in the online replication package (file  SonarQube-rules.xlsx). Note that in column ''Type'' MA means Major, Mi means Minor, CR means Critical, and BL means Blocker.

\subsection{RQ1: Which are the most fault-prone SQ-Violations?}

In order to answer this RQ, we first analyzed the importance of the SQ-Violations by means of the most accurate ML technique and then we performed the residual analysis. 

\subsubsection{SQ-Violations Importance Analysis}

\begin{figure}
\centering
\includegraphics[width=0.9\linewidth]{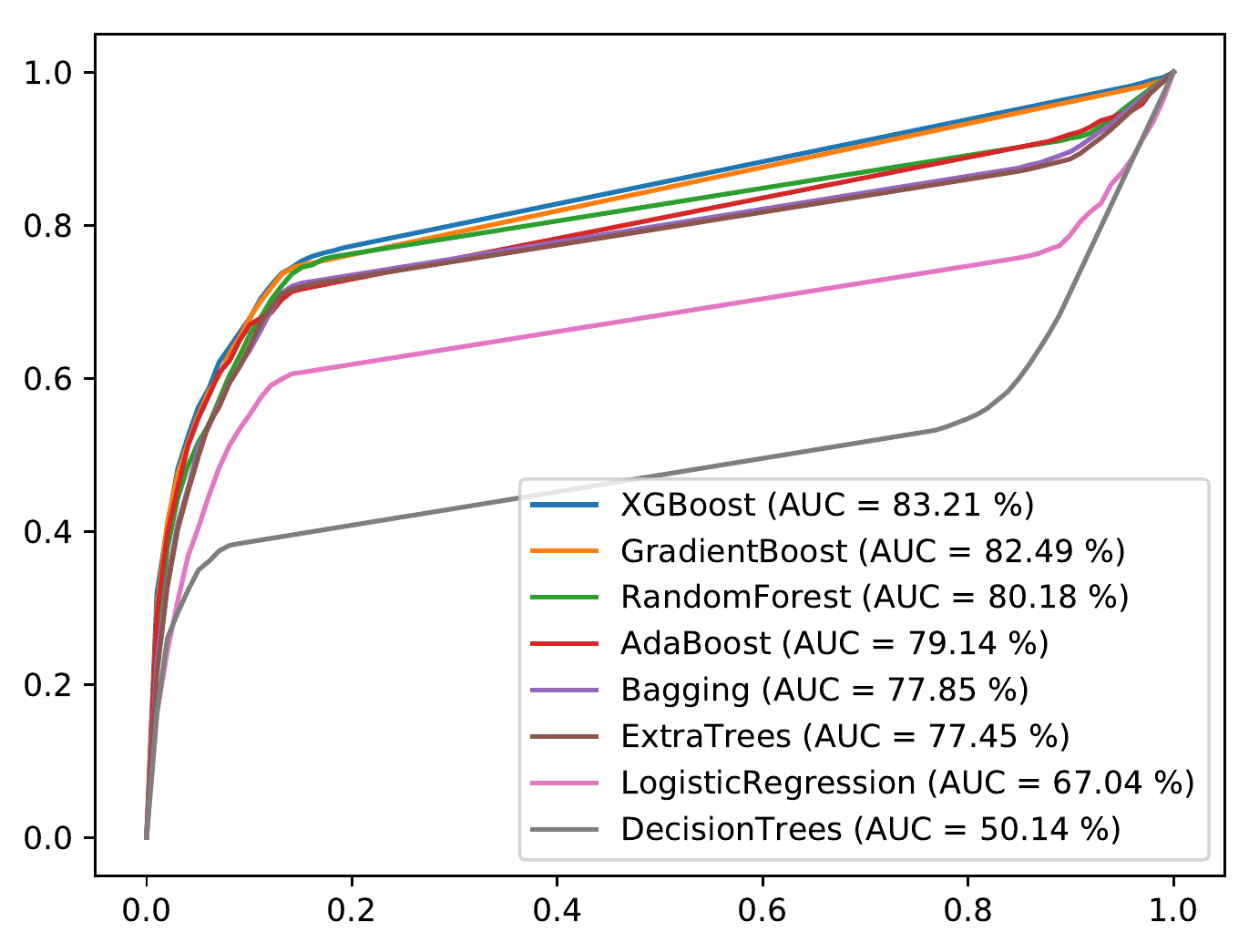}
\vspace{-2mm}
\captionof{figure}{ROC Curve (Average between 10-fold validation models)} \label{fig:ROC_Comparison}
\vspace{-4mm}
\end{figure}

As shown in Figure~\ref{fig:ROC_Comparison}, XGBoost resulted in the most accurate model among the eight machine learning techniques applied to the dataset. The 10-fold cross-validation reported an average AUC of 0.83. Table~\ref{tab:prc} (column RQ1) reports average reliability measures for the eight models. 

Despite the different measures have different strengths and weaknesses (see Section~\ref{sec:accuracy}), all the measures are consistently showing that XGBoost is the most accurate technique.

The ROC curves of all models are depicted in Table~\ref{tab:prc} while the reliability results of all the 10-folds models are available in the online replication package.

%\todo [inline] {here I need help to report the results}

Therefore, we selected XGBoost as classification model for the next steps, and utilized the feature importance calculated applying the drop-column method to this classifier. The XGBoost classifier was retrained removing one feature at a time sequentially. 

% \begin{table*}
% \centering
% \begin{tabular}{ll}

% \begin{minipage}{0.48\textwidth}
% \includegraphics[width=1\linewidth]{./Figure/Squids1.pdf}
% \captionof{figure}{ROC Curve (Average between 10-fold validation models)} \label{fig:ROC_Comparison}
% \end{minipage}

% & 

% \begin{minipage}{0.48\textwidth}
% \includegraphics[width=1.016\linewidth]{./Figure/SonarQubeAUC.pdf}
% \captionof{figure}{ROC Curve of the fault-proneness of SonarQube  violations classified as ''bugs'' } \label{fig:ROC_SQ}
% \end{minipage}

% \end{tabular}
% \end{table*}

% \todo [inline] {Values are related to the old paper. Need to be fixed! List of metrics should also be fixed}
%\todo[inline]{Update new values for Bagging as soon as the script ends running both in Figure 5 and table 3 (and Figure 10 at the end)}
\begin{table*}[!ht]
\centering
\footnotesize
\caption{Model Reliability}
\label{tab:prc}
\begin{tabular}{p{1.3cm}|p{0.9cm}|p{1cm}|p{1cm}|p{1cm}|p{0.8cm}|p{1.3cm}|p{1.2cm}|p{1.2cm}|p{1.3cm}}
\hline
& \multicolumn{7}{c|}{\textbf{RQ1} (Average between 10-fold validation models)} & \textbf{RQ2} & \textbf{RQ3}\\ \hline
\multirow{2}{*}{\textbf{Measure}} & \textbf{Logistic Regr.}& \textbf{Decision Tree}& \textbf{Bagging}& \textbf{Random Forest}& \textbf{Extra Trees}& \textbf{AdaBoost}  & \textbf{Gradient Boosting} & \textbf{XGBoost} & \textbf{SQ ''bugs''}\\
\hline
Precision &0.417 &0.311 &0.404 &0.532 &0.427 &0.481 &0.516 &0.608 &0.086 \\ \hline
Recall &0.076&0.245 &0.220&0.156&0.113&0.232&0.192&0.182&0.028 \\ \hline
MCC &0.162&0.253& 0.279&0.266&0.203&0.319&0.300&0.318&0.032 \\ \hline
f-measure & 0.123&0.266 &0.277&0.228&0.172&0.301&0.275&0.275&0.042 \\ \hline
TNR &0.996&0.983& 0.990&0.995&0.995&0.993&0.995&0.997&0.991\\ \hline
FPR &0.004&0.002 &0.010&0.004&0.005&0.007&0.005&0.003&0.009\\ \hline
FNR &0.924&0.755 &0.779&0.844&0.887&0.768&0.808&0.818&0.972\\ \hline
AUC &0.670&0.501 &0.779&0.802&0.775&0.791&0.825&0.832&0.509\\ \hline
\end{tabular}
\vspace{-5mm}
\end{table*}

% \begin{table}[!h]
% \centering
% \caption{XGBoost Reliability (Average between 5-fold validation models)}
% \vspace{-2mm}
% \label{tab:prc}
% \begin{tabular}{p{1.4cm}|p{1cm}|p{1cm}|p{1cm}|p{1cm}|p{1cm}}
% \textbf{Accuracy Measure} & \textbf{\nth{1}fold}& \textbf{\nth{2}fold}& \textbf{\nth{3}fold}& \textbf{\nth{4}fold}& \textbf{\nth{5}fold} \\
% \hline
% Precision &0.706 &0.379 &0.999 &0.619 &0.997 \\ \hline
% RECALL &0.991&0.988&0.968&0.889&0.988 \\ \hline
% MCC &0.833&0.605&0.715&0.736&0.736 \\ \hline
% f-measure & 0.825&0.751&0.823&0.730&0.843 \\ \hline
% \end{tabular}
% \end{table}

%\todo [inline]{Fix Table with SQ model reliability and add ROC Curve for SQ Model}

% % Please add the following required packages to your document preamble:
% % \usepackage{multirow}
% \begin{table}[!h]
% \centering
% \caption{Contingency Matrix (Average between 5-fold validation models)}
% \label{tab:ourConfusion}
% \begin{tabular}{c c|c|c}
% & & \multicolumn{2}{c}{\textbf{Actual}} \\ \cline{3-4} 
% & \textbf{Predicted}  & IND  & NOT IND  \\ \hline
% & IND & 817  & 339  \\ \hline
% & NOT IND  & 7 & 38355 \\ \hline
% \end{tabular}
% \end{table}

23 SQ-Violations have been ranked with an importance higher than zero by the XGBoost.  In Table~\ref{tab:results}, we report the SQ-Violations with an importance higher or equal than 0.01 \% (coloumn ''Intr. \& Rem. (\%)'' reports the number of violations introduced in the fault-inducing commits AND removed in the fault-fixing commits). The remaining SQ-Violations are reported in the raw data for reasons of space. coloumn ''Intr. \& Rem. (\%)'' means  

The combination of the 23 violations guarantees a good classification power, as reported by the AUC of 0.83. However, the drop column algorithm demonstrates that SQ-Violations have a very low individual importance. The most important SQ-Violation has an importance of 0.62\%. This means that the removal of this variable from the model would decrease the accuracy (AUC) only by 0.62\%. 
Other three violations have a similar importance (higher than 0.5\%) while others are slightly lower.

\subsubsection{Model Accuracy Validation}

The analysis of residuals shows that several SQ-Violations are introduced in fault-inducing commits in more than 50\% of cases. 
32 SQ-Violations out of 202  had been introduced in the fault-inducing commits and then removed in the fault-fixing commit in more than 95\% of the faults. The application of the XGBoost, also confirmed an importance higher than zero in 26 of these SQ-Violations. 
This confirms that developers, even if not using SonarQube, pay attention to these 32 rules, especially in case of refactoring or bug-fixing.

Table~\ref{tab:results} reports the descriptive statistics of residuals, together with the percentage residuals = 0 (number  of SQ-Violations introduced during fault-inducing commits and 
\begin{figure}[!h]
\centering
\includegraphics[trim = 2 70 2 2, clip, width=\linewidth]{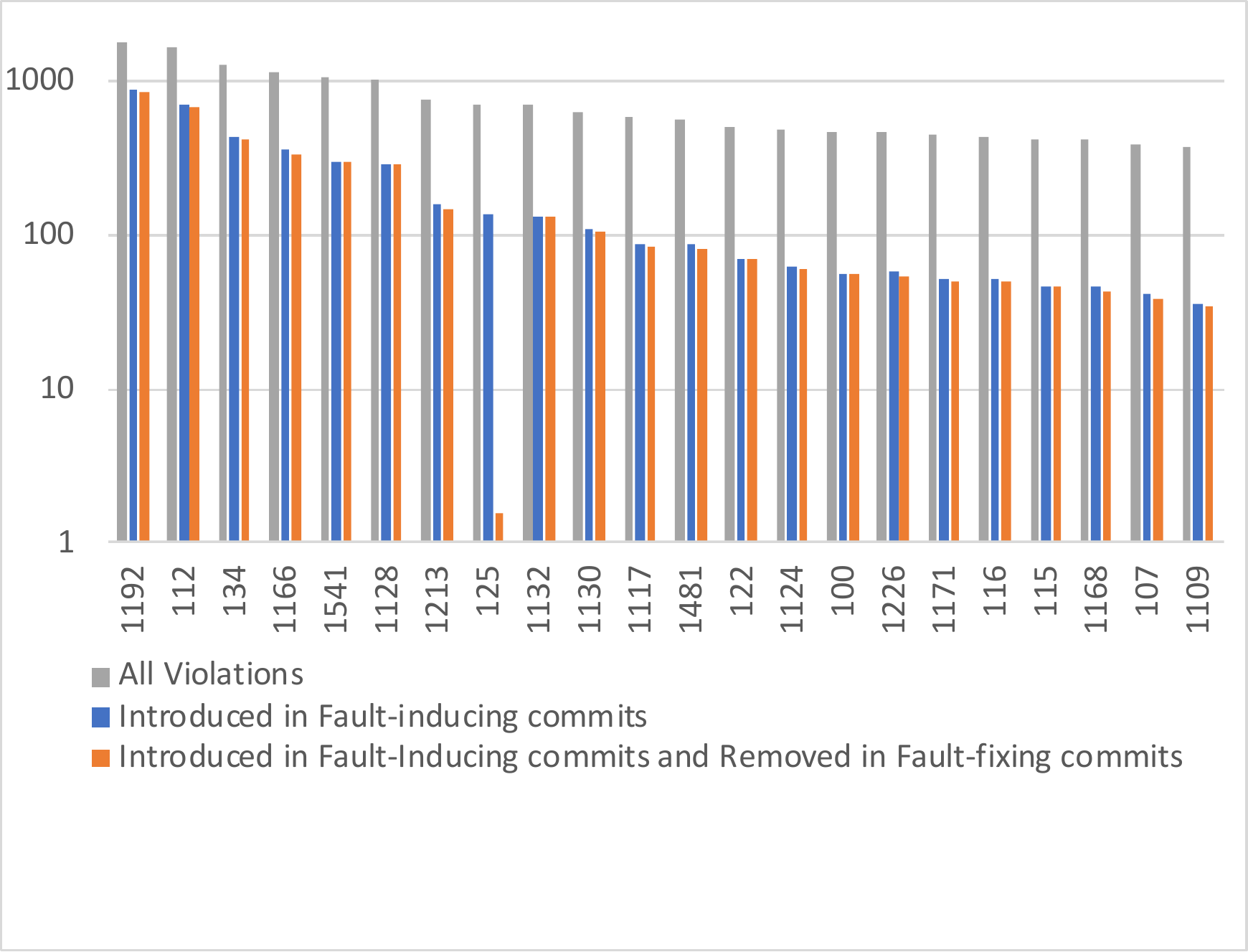}
\vspace{-2mm}
\caption{Comparison of Violations introduced in fault-inducing commits and removed in fault-fixing commits}
\label{fig:imp}
\vspace{-4mm}
\end{figure}

\noindent removed during fault-fixing commits). 

Column "Res $>$95\%", shows a checkmark (\checkmark) when the percentage of residuals=0 was higher than 95\%.

Figure~\ref{fig:imp} compares the number of violations introduced in fault-inducing commits, and the number of violations removed in the fault-fixing commits. 

% \begin{figure*}[!h]
% \centering
% \includegraphics[trim = 0 0 0 0, clip, width=\linewidth]{./Figure/Figure7.png}
% \caption{Comparison of Violations introduced in fault-inducing commits and removed in fault-fixing commits}
% \label{fig:imp}
% \end{figure*}

\subsection{Manual Validation of the Results}
In order to understand the possible causes and to validate the results, we manually analyzed 10 randomly selected instances for the first 20 SQ-Violations ranked as more important by the XGBoost algorithm.

The first immediate result is that, in 167 of the 200 manually inspected violations, the bug induced in the fault-inducing commit  was not fixed by the same developer that induced it. 

We also noticed that violations related to duplicated code and empty statements (eg. "method should not be empty") always generated a fault (in the randomly selected cases). When committing an empty method (often containing only  a "TODO" note), developers often forgot to implement it and then used it without realizing that the method did not return the expected value. An extensive application of unit testing could definitely reduce this issue. However, we are aware that is is a very common practice in several projects. 
Moreover,  SQ-Violations such as 1481 (unused private variable should be removed) and 1144 (unused private methods should be removed) unexpectedly resulted to be an issue. In several cases, we discovered methods not used, but expected to be used in other methods, resulted in a fault. As example, if a method A calls another method B to compose a result message, not calling the method B results in the loss of the information provided by B.  

\subsection{RQ2: Are SQ-Violations classified as ''bugs'' by SonarQube more fault-prone than other rules?}
Out of the 57 violations classified as ''bugs'' by SonarQube, only  three (squid 1143, 1147, 1764) were considered fault-prone with a very low importance from the XGBoost and with residuals higher than 95\%.  However, rules classified as ''code smells'' were frequently violated in fault-inducing commits. 
However, considering all the SQ-Violations, out of 40 the SQ-Violations that we identified as fault-prone, 37 are classified as ''code smells'' and one as security ''vulnerability''.  
% \textbf{ 37 out of 40 fault rules are classified by SonarQube as "code smells", one is "security vulnerabilities". } 
% \todo [inline] {verify these values based on the application of the log. reg.}

When comparing severity with fault proneness of the SQ-Violations, only three SQ-Violations (squid 1147, 2068, 2178) were associated with the highest severity level (blocker). However, the fault-proneness of this rule is extremely low (importance $<$= 0.14\%). Looking at the remaining violations, we can see that the severity level is not related to the importance reported by the XGBoost algorithm since the rules of different level of severity are distributed homogeneously across all importance levels.

\subsection{RQ3: Fault prediction accuracy of the SonarQube model}
''Bug'' violations were introduced in 374 commits out of 39,518 analyzed commits.   Therefore, we analyzed which of these commits were actually fault-inducing  commits. Based on SonarQube's statement, all these commits should have generated a fault.

All the accuracy measures (Table~\ref{tab:prc}, column ''RQ2'') confirm the very low prediction power of ''bug'' violations. The vast majority of ''bug'' violations never become a fault. Results are also confirmed by the extremely low AUC  (50.95\%) and by the contingency matrix (Table~\ref{tab:sq-reliability}). 
The results of the SonarQube model also confirm the results obtained in RQ2. Violations classified as "bugs" should be classified differently since they are hardly ever injected in  fault-inducing commits. 
\begin{table}[]
\centering
\caption{SonarQube Contingency Matrix (Prediction model based on SQ-Violations considered as ''Bug'' by SonarQube)}
\label{tab:sq-reliability}
\begin{tabular}{c c|c|c}
\hline 
& \textbf{Predicted} & \multicolumn{2}{c}{\textbf{Actual}} \\ \cline{3-4} 
&  & IND  & NOT IND  \\ \hline
& IND & 32  & 342  \\ \hline
& NOT IND  & 1,124 & 38,020  \\ \hline
\end{tabular}
\end{table}

\begin{table*}[]
\footnotesize
\caption{Summary of the most important SonarQube Violations Related to Faults (XGBoost Importance $>$ 0.2\%)}
\label{tab:results}
\begin{tabular}{p{2.7cm}|p{1.3cm}|p{0.7cm}|p{0.8cm}|p{1.cm}|p{1cm}|p{0.7cm}|p{0.6cm}|p{0.45cm}|p{0.7cm}|p{0.45cm}|p{0.6cm}|p{0.8cm}} \hline 
\multicolumn{4}{c|}{\textbf{SonarQube}} & \multicolumn{2}{c|}{\textbf{SZZ}} & \multicolumn{5}{c|}{\textbf{Residuals}} & \textbf{XG Boost} & \textbf{Res. $>$95\%} \\ \cline{1-12}  
SQUID	&	Severity	&	Type	&	\# Occ.	&	Intr. \& Rem.(\%) 	&	Intr. in fault-ind	&	Mean	&	Max	&	Min	&	Stdev	&	RSS	&	Imp.	&		\\  \hline 

 S1192	&	CRITICAL	&	CS	&	1815	&	50,87	&	95,10	&	245,60	&	-861	&	2139	&	344,42	&	1726	&	0,66	&	\checkmark	\\ \hline
 S1444	&	MINOR	&	CS	&	96	&	2,69	&	97,92	&	4,59	&	-7	&	73	&	10,34	&	94	&	0,62	&	\checkmark	\\ \hline
 Useless Import Check	&	MAJOR	&	CS	&	1026	&	28,76	&	97,27	&	33,37	&	-170	&	351	&	61,58	&	998	&	0,41	&	\checkmark	\\ \hline
 S00105	&	MINOR	&	CS	&	263	&	7,37	&	97,72	&	1,96	&	-13	&	32	&	10,22	&	257	&	0,41	&	\checkmark	\\ \hline
 S1481	&	MINOR	&	CS	&	568	&	15,92	&	95,25	&	10,41	&	-6	&	83	&	14,60	&	541	&	0,39	&	\checkmark	\\ \hline
 S1181	&	MAJOR	&	CS	&	200	&	5,61	&	97,00	&	8,87	&	0	&	88	&	13,43	&	194	&	0,31	&	\checkmark	\\ \hline
 S00112	&	MAJOR	&	CS	&	1644	&	46,08	&	94,77	&	188,26	&	-279	&	1529	&	270,34	&	1558	&	0,29	&		\\ \hline
 S1132	&	MINOR	&	CS	&	704	&	19,73	&	93,75	&	121,75	&	-170	&	694	&	134,91	&	660	&	0,24	&		\\ \hline
 Hidden Field	&	MAJOR	&	CS	&	584	&	16,37	&	92,98	&	26,96	&	-12	&	143	&	29,42	&	543	&	0,23	&		\\ \hline
 S134	&	CRITICAL	&	CS	&	1272	&	35,65	&	94,65	&	70,66	&	-66	&	567	&	88,07	&	1204	&	0,20	&		\\ \hline
\end{tabular}
 \vspace{-3mm}
\end{table*}

\section{Threats to Validity}
\label{Threat}
In this Section, we discuss the threats to validity, including internal, external, construct validity, and reliability. 
We also explain the different adopted tactics~\cite{YinCaseStudies2009}. 

\textbf{Construct Validity}. As for construct validity, the results might be biased regarding the mapping between faults and commits. We relied on the ASF practice of tagging commits with the issue ID. However, in some cases, developers could have tagged a commit differently. Moreover, the results could also be biased due to detection errors of SonarQube.
\review{
We are aware that static analysis tools suffer from false positives. In this work we aimed at understanding the fault proneness of the rules adopted by the tools without modifying them, so as to reflect the real impact that developers would have while using the tools. In future works, we are planning to  replicate this work manually validating a statistically significant sample of violations, to assess the impact of false positives on the achieved findings. }
As for the analysis timeframe, 
we analyzed commits until the end of 2015, considering all the faults raised until the end of March 2018. We expect that the vast majority of the faults should have been fixed. However, it could be possible that some of these faults were still not identified and fixed.

\textbf{Internal Validity.} Threats can be related to the causation between SQ-Violations and fault-fixing activities. As for the identification of the fault-inducing commits, we relied on the SZZ algorithm~\cite{SZZ}.  We are aware that in some cases, the SZZ algorithm might not have identified fault-inducing commits correctly because of the limitations of the line-based diff provided by git, and also because  in some cases bugs can be fixed modifying code in other location than in the lines that induced them. Moreover, we are aware that the imbalanced data could have influenced the results (approximately 90\% of the commits were non-fault-inducing). However, the application of solid machine learning techniques, commonly applied with imbalanced data could help to reduce this threat.

%  \todo [inline] {Could the qualitative nature of that data be clarified? / Outi}
\textbf{External Validity.} We selected 21 projects from the ASF, which incubates only certain systems that follow specific and strict quality rules. Our case study was not based only on one application domain. This was avoided since we aimed to find general mathematical models for the prediction of the number of bugs in a system. Choosing only one or a very small number of application domains could have been an indication of the non-generality of our study, as only prediction models from the selected application domain would have been chosen. The selected projects stem from a very large set of application domains, ranging from external libraries, frameworks, and web utilities to large computational infrastructures. 
% The application domain was not an important criterion for the selection of the projects to be analyzed, but in any case we tried to balance the selection and pick systems from as many contexts as possible.
\review{The dataset only included Java projects. We are aware that different  programming languages, and projects different maturity levels could provide different results.} 

\textbf{Reliability Validity.} We do not exclude the possibility that other statistical or machine learning approaches such as  Deep Learning, or others might have yielded similar or even better accuracy than our modeling approach.

\section{Related Work}
\label{relworks}
In this Section, we introduced the related works analyzing literature on SQ-Violations and faults predictions. 

Falessi et al.~\cite{Falessi2017} studied the distribution of  16 metrics and 106 SQ-Violations in an industrial project. They applied a \textit{What-if} approach with the goal of investigating what could happen if a specific SQ-Violation would not have been introduced in the code and if the number of faulty classes  decrease in case the violation is not introduced. They compared  four ML techniques  applying the same techniques on a  modified version of the code where they manually removed SQ-Violations. Results showed that 20\% of faults were avoidable if the code smells would have been removed. 

Tollin et al.~\cite{Tollin2017} investigated if SQ-Violations introduced would led to an increase in the number of changes (code churns) in the next commits. The study was applied on two different industrial projects, written in C\# and JavaScript. 
% They compared the prediction accuracy of Decision Trees, Random Forest, and Naive Bayes.  
They reported that classes affected by more SQ-Violations have a higher change proneness.  However they did not prioritize or classified the most change prone SQ-Violations.

Digkas et al.~\cite{digkas2018} studied weekly snapshots of 57  Java projects of the ASF
% with the following requirements: the projects had to be written in Java programming language, had to have at least 100 classes, at least two years of history, at least 1,000 commits and had to be still active in the year the project was conducted. Their goal was to find out how much 
investigating the amount of technical debt  paid back over the course of the projects and what kind of issues were fixed. They considered SQ-Violations with severity marked as \textit{Blocker, Critical, and  Major}.
The results showed that only a small subset of all issue types was responsible for the largest percentage of technical debt repayment. Their results thus confirm our initial assumption that there is no need to fix all issues. Rather, by targeting particular violations, the development team can achieve higher benefits. However, their work does not consider how the issues actually related to faults.

Falessi and Reichel~\cite{falessi-reichel} developed an open-source tool to analyze the technical debt interest occurring due to violations of quality rules. Interest is measured by means of various metrics related to fault-proneness. They use SonarQube rules and uses linear regression to estimate the defect-proneness of classes. The aim of MIND is to answer  developers' questions like: is it worth to re-factor this piece of code? Differently  than in our work, the actual type of issue causing the defect was not considered.

Codabux and Williams~\cite{Codabux-williams} propose a predictive model to  prioritize technical debt. They extracted class-level metrics for defect- and change-prone classes using \textit{Scitool Understanding} and  \textit{Jira Extracting Tool} from Apache Hive and determined significant independent variables for defect- and change-prone classes, respectively. Then they used a Bayesian approach to build a prediction model to determine the ''technical debt proneness'' of each class. Their model requires the identification of ''technical debt items'', which requires manual input. These items are ultimately ranked and given a risk probability by the predictive framework.

Saarimäki investigated the diffuseness of SQ-violations in the same dataset we adopted~\cite{Saarimaki2019} and the accuracy of the SonarQube remediation time~\cite{SaarimakiEUROMICRO2019}.

Regarding other code quality rules detection, 7 different machine learning approaches (Random Forest, Naive Bayes, Logistic regression, IBl, IBk, VFI, and J48)~\cite{Maneerat2011} were successfully applied on 6 code smells (Lazy Class, Feature Envy, Middle Man Message Chains, Long Method, Long Parameter Lists, and Switch Statement) and 27 software metrics (including Basic, Class Employment, Complexity, Diagrams, Inheritance, and MOOD) as independent variables. 

Code smells detection was also investigated from the point of view of how the severity of code smells can be classified through machined learning models~\cite{Fontana2017} such as J48, JRip, Random Forest, Naive Bayes, SMO, and LibSVM with best agreement to detection 3 code smells (God Class, Large Class, and Long Parameter List).

\begin{table*}[H]
\footnotesize
\caption{Summary of the SonarQube Violations Related to Faults (XGBoost Importance $>$ 0.03\%)}
\label{tab:results}
\begin{tabular}{p{1.2cm}|c|p{0.7cm}|p{0.6cm}|p{0.9cm}|p{0.65cm}|p{0.65cm}|p{0.6cm}|p{0.45cm}|p{0.7cm}|p{0.45cm}|p{0.6cm}|p{0.5cm}} \hline 
\multicolumn{4}{c|}{\textbf{SonarQube}} & \multicolumn{2}{c|}{\textbf{SZZ}} & \multicolumn{5}{c|}{\textbf{Residuals}} & \textbf{XG Boost} & \textbf{Res.$>$ 95\%} \\ \cline{1-12}  
SQUID	&	Severity	&	Type	&	\# Occ.	&	Intr. \& Rem. (\%) 	&	Intr. in fault-ind	&	Mean	&	Max	&	Min	&	Stdev	&	RSS	&	Imp.	&		\\  \hline 
 S1192	&	CRITICAL	&	CS	&	1815	&	50.87	&	95.10	&	245.60	&	-861	&	2139	&	344.42	&	1726	&	0.66	&	\checkmark	\\ \hline
 S1444	&	MINOR	&	CS	&	96	&	2.69	&	97.92	&	4.59	&	-7	&	73	&	10.34	&	94	&	0.62	&	\checkmark	\\ \hline
 Useless Import Check	&	MAJOR	&	CS	&	1026	&	28.76	&	97.27	&	33.37	&	-170	&	351	&	61.58	&	998	&	0.41	&	\checkmark	\\ \hline
 S00105	&	MINOR	&	CS	&	263	&	7.37	&	97.72	&	1.96	&	-13	&	32	&	10.22	&	257	&	0.41	&	\checkmark	\\ \hline
 S1481	&	MINOR	&	CS	&	568	&	15.92	&	95.25	&	10.41	&	-6	&	83	&	14.60	&	541	&	0.39	&	\checkmark	\\ \hline
 S1181	&	MAJOR	&	CS	&	200	&	5.61	&	97.00	&	8.87	&	0	&	88	&	13.43	&	194	&	0.31	&	\checkmark	\\ \hline
 S00112	&	MAJOR	&	CS	&	1644	&	46.08	&	94.77	&	188.26	&	-279	&	1529	&	270.34	&	1558	&	0.29	&		\\ \hline
 S1132	&	MINOR	&	CS	&	704	&	19.73	&	93.75	&	121.75	&	-170	&	694	&	134.91	&	660	&	0.24	&		\\ \hline
 Hidden Field	&	MAJOR	&	CS	&	584	&	16.37	&	92.98	&	26.96	&	-12	&	143	&	29.42	&	543	&	0.23	&		\\ \hline
 S134	&	CRITICAL	&	CS	&	1272	&	35.65	&	94.65	&	70.66	&	-66	&	567	&	88.07	&	1204	&	0.20	&		\\ \hline
 S1068	&	MAJOR	&	CS	&	471	&	13.20	&	97.24	&	7.07	&	-39	&	77	&	13.17	&	458	&	0.17	&	\checkmark	\\ \hline
 S1186	&	CRITICAL	&	CS	&	369	&	10.34	&	94.85	&	12.72	&	-7	&	64	&	12.77	&	350	&	0.17	&		\\ \hline
 S106	&	MAJOR	&	CS	&	267	&	7.48	&	92.51	&	7.25	&	-172	&	106	&	38.13	&	247	&	0.16	&		\\ \hline
 S00108	&	MAJOR	&	CS	&	302	&	8.46	&	94.04	&	18.54	&	-19	&	149	&	31.06	&	284	&	0.16	&		\\ \hline
 Redundant Throws Declaration	&	MAJOR	&	CS	&	639	&	17.91	&	94.84	&	93.28	&	-265	&	593	&	114.34	&	606	&	0.15	&		\\ \hline
 S1147	&	BLOCKER	&	BUG	&	35	&	0.98	&	100.00	&	29.23	&	4	&	66	&	24.86	&	35	&	0.14	&	\checkmark	\\ \hline
 S00119	&	MAJOR	&	CS	&	24	&	0.67	&	0.00	&	26.00	&	26	&	26	&	0.00	&	0	&	0.14	&	\checkmark	\\ \hline
 S1172	&	MAJOR	&	CS	&	272	&	7.62	&	98.90	&	8.13	&	-3	&	101	&	11.81	&	269	&	0.13	&	\checkmark	\\ \hline
 S00115	&	MINOR	&	CS	&	419	&	11.74	&	95.47	&	31.81	&	-53	&	166	&	36.21	&	400	&	0.13	&	\checkmark	\\ \hline
 S00116	&	MAJOR	&	CS	&	433	&	12.14	&	97.46	&	37.49	&	-1681	&	1881	&	377.83	&	422	&	0.13	&	\checkmark	\\ \hline
 S1199	&	MINOR	&	CS	&	107	&	3.00	&	96.26	&	11.53	&	-223	&	259	&	51.37	&	103	&	0.12	&	\checkmark	\\ \hline
\end{tabular}
\end{table*}

\section{Discussion and Conclusion}
\label{Conclusions}
SonarQube classifies 57 rules as "bugs", claiming that they will sooner or later they generate faults. Four local companies contacted us to investigate the fault prediction power of the SonarQube rules, possibly using machine learning, so as to understand if they can rely on the SonarQube default rule-set or if they can use machine learning to customize the model more accurately. 
% Therefore,  pro In this work, we performed a large case study with the goal of analyzing the fault-proneness of SonarQube violations and the prediction power of the default SonarQube quality model used by SonarQube to calculate the technical debt of Java code. 
% The SonarQube model is composed of 202 rules that, when violated, generate  issues that should be addressed (removed from the code) by the developers. 

% Moreover, among these rules, 
We conducted this work analyzing a set of 21 well-known open source project selected by the companies, analyzing the presence of all 202 SonarQube detected violations in the complete project history. 
% analyzing all the commits from the beginning of the projects until the end of 2015. We identified and labeled the fault-inducing commits, mapping the faults reported in the Jira issue tracker by means of the SZZ algorithm. 
The study considered 39,518 commits, including more than 38 billion lines of code, 1.4 million violations, and 4,505 faults mapped to the commits.

To understand which sq-violations have the highest fault-proneness, we first applied eight machine learning approaches to identify the sq-violations that are common in commits labeled as fault-inducing. As for the application of the different machine learning approaches, we can see an important difference in their accuracy, with a difference of more than 53\% from the worst model (Decision Trees AUC=47.3\%$\pm$3\%) and the best model (XGBoost AUC=83.32\%$\pm$10\%). This confirms also what we reported in Section~\ref{MachineLearning}: ensemble models, like the XGBoost, can generalize better the data compared to Decision Trees, hence it results to be more scalable. The use of many \textit{weak} classifiers, yields an overall better accuracy, as it can be seen by the fact that the \textit{boosting} algorithms (AdaBoost, GradientBoost, and XGBoost) are the best performers for this classification task, followed shortly by the Random Forest classifier and the ExtraTrees. %\todo [inline]{Add sth to motivate if, if you think is useful} 

As next step, we checked the percentage of commits where a specific violation was introduced in the fault-inducing commit and then removed in the fault-fixing commit, accepting only those violations where the percentage of cases where the same violations were added in the fault-inducing commit and removed in the fault-fixing commit was higher than 95\%.

Our results show that 26 violations can be considered fault-prone from the XGBoost model. However, the analysis of the residuals showed that 32 sq-violations were commonly introduced in a fault-inducing commit and then removed in the fault-fixing commit but only two of them are considered fault-prone from the machine learning algorithms. 
 It is important to notice that all the sq-violations that are removed in more than 95\% of cases during fault-fixing commits are also selected by the XGBoost, also confirming the importance of them. 

When we looked at which of the sq-violations were considered as fault-prone in the previous step, only four of them are also classified as (''bugs'') by SonarQube. The remaining  fault-prone sq-violations are mainly classified as ''code smells'' (SonarQube claims that ''code smells'' increase maintenance effort but do not create faults). 
The analysis of the accuracy of the fault prediction power of the SonarQube model based on ''bugs'' showed an extremely low fitness, with an AUC of 50.94\%, confirming that violations classified as ''bugs'' almost never resulted in a fault.   

An important outcome is related to the application of the machine learning techniques. Not all the techniques performed equally and XGBoost was the most more accurate and fastest technique in all the projects. Therefore, the application XGBoost to historical data is a good alternative to the manual tuning of the model, where developers should select which rules they believe are important based on their experience. 

\review{The result confirmed the impression of the developers of our companies. Their developers still consider it very useful to help to develop clean code that adhere to company standards, and that help new developers to write code that can be easily understood by other developers. Before the execution of this study the companies were trying to avoid to violate the rules classifies as bugs, hoping to reduce fault proneness. However, after the execution of this study, the companies individually customized the set of rules considering only coding standards aspects and rules classified as ''security vulnerabilities''. The main result for the companies is that they will need to invest in the adoption of other tools to reduce the fault proneness and therefore, we will need to replicate this work considering other tools such as FindBugs, PMD but also commercial tools such as Coverity Scan, Cast Software and others.}

%  Our findings clearly show that at least 32 sq-violations should be carefully monitored by developers, since all of them are related to faults. Table~\ref{tab:results} shows a summary of our findings, where we ranked the sq-violations based on their fault-proneness and we reported how often they are introduces in fault-inducing commits and removed in fault-fixing commits. Looking at the table we can see that the removal of the 32 sq-violations that are considered important from the XGBoost algorithm and also have zero residuals in more than 95\% of cases could provide a high benefit in terms of fault -proneness: thus, on the one hand practitioners should carefully monitor these sq-violations and plan refactoring actions to improve the overall maintainability of the code.

Based on the overall results, we can summarize the following lessons learned:

\textbf{Lesson 1:} SonarQube violations are not good predictors of fault-proneness if considered individually, but can be good predictors if considered together.  Machine learning techniques, such as XGBoost can be used to effectively train a customized model for each company. 

\textbf{Lesson 2:} SonarQube violations classified as  ''bugs'' do not seem to be the cause of faults. 

\textbf{Lesson 3:} SonarQube violation severity is not related to the fault-proneness and therefore, developers should carefully consider the severity as decision factor for refactoring a violation. 

\textbf{Lesson 4:} Technical debt should be calculated differently, and the non-fault prone rules should not be accounted as ''fault-prone'' (or ''buggy'') components of the technical debt while several ''code smells'' rules should be carefully considered as potentially fault-prone.

The lessons learned confirm our initial hypothesis about the fault-proneness of the SonarQube violations. However,  we are not claiming that SonarQube violations are not harmful in general. We are aware that some violations could be more prone to changes~\cite{Falessi2017}, decrease code readability, or increase the maintenance effort.  

Our recommendation to companies using SonarQube is to customize the rule-set, taking into account which violations to consider, since the refactoring of several sq-violations might not lead to a reduction in the number of faults. Furthermore, since the rules in SonarQube constantly evolve, companies should continuously re-consider the adopted rules.  

Research on technical debt should focus more on validating which rules are actually harmful from different points of view and which will account for a higher technical debt if not refactored immediately.

Future works  include the replication of this work considering the   severity levels of SonarQube rules and their importance. We are working on the definition of a more accurate model for predicting TD~\cite{LenarduzziMALTESQUE2019}
% We are currently working on an analysis of all the commits of the Java projects in the Apache Software Foundation repository.
 Moreover,  \review{ we
 are planning to investigate whether classes that SonarQube  identify as problematic are more fault-prone than those not affected by any problem. Since this work did not confirmed the fault proneness of SonarQube rules, the companies are interested in finding other static analysis tool for this purpose. Therefore,  we are planning to replicate this study using other tools such as FindBugs, Checkstyle, PMD and others.}
 Moreover, we will focus on the definition of recommender systems integrated in the IDEs~\cite{Janes2017}\cite{LenarduzziQCM2017},
 to alert developers about the presence of potential problematic classes based on their (evolution of) change- and fault-proneness and rank them based on the potential benefits provided by their removal.

\bibliographystyle{unsrt}
\bibliography{sample-bibliography}

\end{document}